\newcommand{\ourtitle}{Effectiveness of Anonymization in Double-Blind Review}
\newcommand{\pdftitle}{\ourtitle}
\newcommand{\pdfauthors}{Claire Le Goues, Yuriy Brun, Sven Apel, Emery Berger, Sarfraz Khurshid, and Yannis Smaragdakis}

\documentclass[sigconf]{acmart}

\makeatletter

\let\proof\@undefined
\let\endproof\@undefined
\makeatother

\usepackage{amsmath}
\usepackage{amsfonts}
\usepackage{amssymb}
\usepackage{amsthm}
\usepackage{balance}
\usepackage{graphicx}
\usepackage{xspace}
\usepackage{pslatex}
\usepackage{pifont}
\usepackage{multirow}
\usepackage{array}
\usepackage{booktabs}
\usepackage{cancel}
\usepackage{microtype}
\usepackage{enumitem}
\usepackage{color,colortbl}
\usepackage{dblfloatfix}
\usepackage{fixltx2e}
\usepackage{subfig}
\usepackage[font=small]{caption}

\usepackage{listings}

\newcommand\lt[1]{{\lstinline+#1+}} 
\renewcommand\t[1]{{\lstinline+#1+}} 
\definecolor{dkgreen}{rgb}{0,0.5,0}
\definecolor{dkred}{rgb}{0.5,0,0}
\definecolor{gray}{rgb}{0.5,0.5,0.5}
\let\origthelstnumber\thelstnumber
\makeatletter
\newcommand*\Suppressnumber{%
  \lst@AddToHook{OnNewLine}{%
    \let\thelstnumber\relax%
     \advance\c@lstnumber-\@ne\relax%
    }%
}

\newcommand*\Reactivatenumber{%
  \lst@AddToHook{OnNewLine}{%
   \let\thelstnumber\origthelstnumber%
   \advance\c@lstnumber\@ne\relax}%
}

\definecolor{LightGray}{gray}{0.9}
\definecolor{Gray}{gray}{0.8}

\usepackage{tikz}

\usepackage{hyperref}
\hypersetup{%
  bookmarksdepth = {-2},
	hidelinks=true,
  pdftitle = {\pdftitle},
  pdfkeywords = {},
  pdfauthor = {\pdfauthors},
  pdfstartview={FitH},
}

\setcopyright{none}
\acmISBN{978-1-4503-5105-8/17/09}
\acmConference[CACM]{CACM}{Viewpoint article submission}{under review}
\acmYear{2017}

\begin{document}

\title{\ourtitle}

\author{Claire Le~Goues}
      \affiliation{\institution{Carnegie Mellon University}}
      \email{clegoues@cs.cmu.edu}
\author{Yuriy Brun}
      \affiliation{\institution{University of Massachusetts Amherst}}
      \email{brun@cs.umass.edu}
\author{Sven Apel}
      \affiliation{\institution{University of Passau}}
      \email{apel@uni-passau.de}
\author{Emery Berger}
      \affiliation{\institution{University of Massachusetts Amherst}}
      \email{emery@cs.umass.edu}
\author{Sarfraz Khurshid}
      \affiliation{\institution{University of Texas Austin}}
      \email{khurshid@ece.utexas.edu}
\author{Yannis Smaragdakis}
      \affiliation{\institution{University of Athens}}
      \email{smaragd@di.uoa.gr}

\renewcommand{\shortauthors}{Le~Goues et al.}
\renewcommand{\footnotetextcopyrightpermission}[1]{\relax}

\begin{abstract}

\looseness-1
Double-blind review relies on the authors' ability and willingness to
effectively anonymize their submissions. We explore anonymization
effectiveness at ASE 2016, OOPSLA 2016, and PLDI 2016 by asking reviewers if
they can guess author identities. We find that
74\%--90\% of reviews contain no correct guess and that
reviewers who self-identify as experts on a paper's topic are more likely to
attempt to guess, but no more likely to guess
correctly. We present our findings, summarize the PC chairs' comments about
administering double-blind review, discuss the advantages and
disadvantages of revealing author identities part of the way through the
process, and conclude by advocating for the continued use of double-blind
review. 

\medskip
\noindent
\textbf{Accepted for publication}: C.~Le~Goues, Y.~Brun, S.~Apel, E.~Berger,
S.~Khurshid, and Y.~Smaragdakis. Effectiveness of Anonymization in
Double-Blind Review. \emph{Communications of the ACM}, in press.

\end{abstract}

\maketitle

\pagestyle{plain}
\thispagestyle{plain}

\section{Introduction}
\label{sec:Introduction}

Peer review is a cornerstone of the academic publication process but can be
subject to the flaws of the humans who perform it. 
Evidence suggests that subconscious biases influence one's ability to objectively
evaluate work:
In a controlled experiment with two disjoint program committees, the ACM
International Conference on Web Search and Data Mining (WSDM'17) found that
reviewers with author
information were $1.76\times$ more likely to recommend acceptance of papers
from famous authors, and $1.67\times$ more likely recommend acceptance of
papers from top institutions~\cite{Tomkins17}. 
A study of three years of the Evolution of Languages conference (2012, 2014,
and 2016) found that, when
reviewers knew author identities, review scores for papers with male first
authors were 19\% higher, and for papers with female first
authors 4\% lower~\cite{Roberts16}.
In a medical discipline, US reviewers were more likely 
to recommend acceptance of papers from US-based institutions~\cite{Gastroenterology98}.  

These biases can affect anyone,
regardless of the evaluator's race and gender~\cite{Moss-racusin12}.
Luckily,
double-blind
review can mitigate these effects~\cite{Gastroenterology98, Tomkins17, Budden08}
and reduce the perception of bias~\cite{Snodgrass06}, making it
a constructive step toward a
review system that objectively evaluates papers based strictly on
the quality of the work.

Three conferences in software engineering and programming languages held in
2016\,---\,the IEEE/ACM International Conference on Automated Software
Engineering (ASE), ACM International Conference on Object-Oriented
Programming, Systems, Languages, and Applications (OOPSLA), and the ACM
SIGPLAN Conference on Programming Language Design and Implementation
(PLDI)\,---\,collected data on anonymization
effectiveness, which we\footnote{Sven Apel and Sarfraz Khurshid were the ASE'16
PC chairs, Claire Le~Goues and Yuriy Brun were the ASE'16 review process chairs,
Yannis Smaragdakis was the OOPSLA'16 PC chair, and Emery Berger was the
PLDI'16 PC chair.} use to assess the degree to which reviewers were able to
successfully deanonymize the papers' authors.
We find that anonymization is imperfect but fairly effective: 70\%--86\% of
the reviews were submitted with no author guesses, and 74\%--90\%
of reviews were submitted with no correct guesses. Reviewers who believe themselves
to be experts on a paper's topic were more likely to attempt to guess author identities
but no more likely to guess correctly. Overall, we 
strongly support the continued use of double-blind review,
finding the extra administrative effort minimal and
well-worth the benefits. 

\section{Methodology}
\label{sec:methodology}

The authors submitting to ASE 2016, OOPSLA 2016, and PLDI 2016 were
instructed to omit author information from the author block and obscure, to
the best of their ability, identifying information in the paper. PLDI
authors were also instructed not to advertise their work.
ASE desk-rejected submissions that listed author information on the first page,
but not those that inadvertently revealed such information in the text.
Authors of OOPSLA submissions who revealed author identities were
instructed to remove the identities, which they did, and no paper was
desk-rejected for this reason. PLDI desk-rejected submissions that revealed
author identities in any way.

The review forms included optional questions about author identities, the
answers to which were only accessible to the PC chairs. The 
questions asked if the reviewer thought he or she knew the identity of at
least one author,
and if so, to make a guess and to select what informed the guess. 
The data considered here refer to the first submitted version of each review.
For ASE, author identities were revealed to reviewers immediately after
submission of an initial review; for OOPSLA, ahead of the PC meeting; for PLDI,
only for accepted papers, after all acceptance decisions were made.

\textbf{Threats to validity.} Reviewers were urged to provide a guess if they
thought they knew an author. A lack of a guess could signify not following those 
instructions. However, this risk is small, e.g., OOPSLA PC members
were allowed to opt out uniformly and yet 83\% of the PC members participated. 
Asking reviewers if they could guess author identities may have
affected their behavior: they may not have thought
about it had they not been asked. Data about reviewers' confidence in
guesses may affect our conclusions.
Reviewers could submit multiple guesses per paper and be considered
correct if at least one guess matched, so making many uninformed guesses
could be considered correct, but we did not observe this phenomenon. In a form of
selection bias, all conferences' review processes were chaired by, and
this article is written by, researchers who support double-blind
review.

\section{Anonymization Effectiveness}
\label{sec:results}

\begin{figure}[t]
\centering
\small
\begin{tabular}{lccc}
\toprule
                                &         ASE    &        OOPSLA    &           PLDI  \\
\midrule
reviewers                       & \phantom{0,0}79 & \phantom{0,0}37   & \phantom{0,}111 \\
papers accepted                 & \phantom{0,0}71 & \phantom{0,0}52   & \phantom{0,0}48 \\
papers rejected   & \phantom{0,}263          & \phantom{0,}144             & \phantom{0,}240 \\
\midrule

\textbf{reviews}                       &      \textbf{1,029}      &           \textbf{\phantom{0,}636}   &           \textbf{1,154} \\
did not contain a correct author guess &           90.2\% &           74.4\% &           81.0\% \\
did not contain an author guess        &           86.4\% &           70.0\% &           74.3\% \\
tried to guess at least one author     &           14.7\% &           30.0\% &           25.7\% \\
guessed at least one author correctly  & \phantom{0}9.8\% &           25.6\% &           19.1\% \\
all author guesses incorrect           & \phantom{0}3.8\% & \phantom{0}4.4\% & \phantom{0}6.7\% \\
\midrule

\textbf{reviews with a guess}   & \textbf{\phantom{0,}140} &           \textbf{\phantom{0,}191}   & \textbf{\phantom{0,}297} \\ 
guess at least one author correctly   & 72.1\% & 85.3\% &           74.1\% \\
guess all authors incorrectly         & 27.9\% & 14.7\% &           25.9\% \\
\midrule

\textbf{papers reviewed}        & \textbf{\phantom{0,}334} &           \textbf{\phantom{0,}196}   & \textbf{\phantom{0,}288} \\ 
no one tried guessing authors        & 66.5\% &           41.8\% & 40.6\% \\
someone guessed an author correctly  & 24.6\% &           50.0\% & 44.1\% \\
all guesses incorrect                & \phantom{0}9.0\% & \phantom{0}8.2\% & 15.3\% \\

\bottomrule
\end{tabular}

\caption{Papers, reviews, reviewers, and author guesses. Reviewers
include those on the program and external committees, but exclude chairs. All
papers received at least three reviews; review load was non-uniform.}
\label{fig:basicstats}
\end{figure}

For the three conferences, 70\%--86\% of reviews were submitted without
guesses, suggesting that reviewers typically did not believe they knew or
were not concerned with who wrote most of the papers they reviewed.
Figure~\ref{fig:basicstats} summarizes the number of reviewers, papers, and
reviews processed by each conference, and the distributions of author
identity guesses.

\begin{figure}[b]
\centering
\small
\begin{tabular}{l|rr|rr|rr}
\toprule
  & \multicolumn{2}{c|}{ASE} & \multicolumn{2}{c|}{OOPSLA} & \multicolumn{2}{c}{PLDI} \\
  & guess  & correct & guess & correct & guess & correct \\\midrule
X & 19.0\% & 74.7\%  & 33.6\% & 86.7\% & 33.7\% & 74.2\% \\
Y & 11.2\% & 71.2\%  & 29.3\% & 84.3\% & 24.6\% & 69.0\% \\
Z & 7.1\%  & 55.6\%  & 21.2\% & 83.3\% & 19.7\% & 48.6\% \\
\bottomrule
\end{tabular}
\caption{Guess rate, and correct guess rate, by self-reported reviewer expertise
  score (X: expert, Y: knowledgable, Z: informed outsider).  
}
\label{fig:expertise}
\end{figure}

When reviewers did guess, they were more likely to be correct 
(ASE 72\% of guesses were correct, OOPSLA 85\%, and PLDI 74\%).
However, 75\% of ASE, 50\% of OOPSLA, and 44\% of PLDI papers had no
reviewers correctly guess even one author, and most reviews
contained no correct guess (ASE 90\%, OOPSLA 74\%, PLDI 81\%).

\textbf{Are experts more likely to guess and guess correctly?}
All reviews included a self-reported assessment of reviewer
expertise (X for expert, Y for knowledgable, and Z for informed outsider).
Figure~\ref{fig:expertise} summarizes guess incidence and guess
correctness by reviewer expertise. For each conference, X reviewers were
statistically significantly more likely to guess than Y and Z
reviewers ($p \leq 0.05$). But the
differences in guess correctness were not significant, except
the Z reviewers for PLDI were statistically significantly correct less often than
the X and Y reviewers ($p \leq 0.05$).
We conclude that reviewers who considered themselves experts were
more likely to guess author identities, but were
no more likely to guess correctly.

\begin{figure}[t]
\includegraphics[width=\columnwidth]{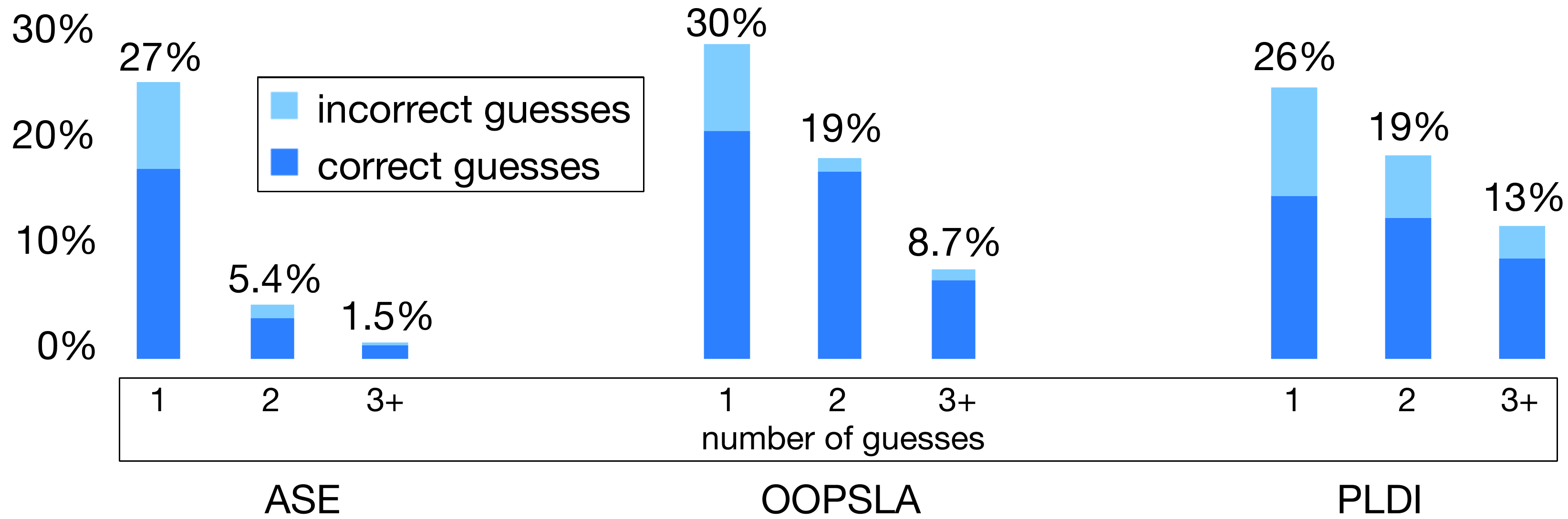}
\caption{Distributions of papers by number of guesses. The
bar shading indicates the fraction of the guesses that are correct.}
  \label{fig:histByPaper}
\end{figure}

\textbf{Are papers frequently poorly anonymized?}  
One possible reason for deanonymization is poor anonymization. Poorly
anonymized papers may have more reviewers guess, and also a higher correct
guess rate.
Figure~\ref{fig:histByPaper} shows the distribution of papers by the number
of reviewers who attempted to guess the authors. The largest proportion
of papers (26\%--30\%) had only a single reviewer attempt to guess. 
Fewer papers had more guesses. The bar shading indicates the
fractions of the author identity guesses that are correct;
papers with more guesses have lower rates of incorrect guesses. Combining
the three conferences' data, the $\chi^2$ statistic indicates that
the rates of correct guessing for papers with one, two, and three or more
guesses are statistically significantly different ($p \leq 0.05$).
This comparison is also statistically significant for OOPSLA alone, but not for 
ASE and PLDI.
Comparing guess rates (we use one-tailed $z$ tests for all population proportion
comparisons)
between paper groups directly: For
OOPSLA, the rate of correct guessing is statistically significantly different
between one-guess papers and each of the other two paper groups. For PLDI,
the same is true between one-guess and three-plus-guess paper groups.
This evidence suggests that a minority of papers may be easy to unblind. 
For ASE, only 1.5\% of the papers had
three or more guesses, while for PLDI, 13\% did. However, for PLDI, 40\% of
all the guesses corresponded to those 13\% of the papers, so
improving the anonymization of a relatively small number of papers
would potentially significantly reduce the number of guesses. Since the three 
conferences only began using the double-blind review process recently, 
the occurrences of insufficient anonymization are likely to decrease
as authors gain more experience with anonymizing submissions, further
increasing double-blind effectiveness.

\begin{figure}[b]
\centering
\small
\begin{tabular}{lccc}
\toprule
papers with        &  ASE    & OOPSLA & PLDI \\
\midrule 
no guesses         &  21.2\% & 20.7\% &  \phantom{0}6.8\% \\
at least one correct guess      &  22.0\% & 31.6\% & 22.3\% \\
all guesses incorrect  &  23.0\% & 25.0\% & 25.0\% \\
\midrule
all papers         &  21.3\% & 26.5\% & 16.7\% \\ 
\bottomrule
\end{tabular}
\caption{Acceptance rate of papers by reviewer guessing behavior. 
}
  \label{fig:acceptance}
\end{figure}

\textbf{Are papers with guessed authors more likely to be accepted?}
We investigated if paper acceptance correlated with either the reviewers'
guesses or with correct guesses. Figure~\ref{fig:acceptance} shows the acceptance rate for each
conference for papers without guesses, with at least one
correct guess, and with all incorrect guesses. We observed different behavior
at the three conferences: ASE submissions were accepted at statistically the same
rate regardless of reviewer guessing behavior.
Additional data available for ASE shows that for each review's
paper rating (strong accept, weak accept, weak reject, strong reject), there
is no statistically significant differences in acceptance rates for submissions
with different guessing behavior. 
OOPSLA and PLDI submissions with no guesses were less likely to be accepted ($p
\leq 0.05$) than those with at least one correct guess. PLDI submissions with no
guesses were also less likely to be accepted ($p \leq 0.05$) than submissions with
all incorrect guesses (for OOPSLA, for the same test, $p = 0.57$).
One possible explanation is that OOPSLA and PLDI reviewers were more likely
to affiliate work they perceived as of higher-quality with known researchers,
and thus more willing to guess the authors of submissions they wanted to
accept.

\textbf{How do reviewers deanonymize?}
OOPSLA and PLDI reviewers were asked if the use of
citations revealed the authors. Of the reviews with guesses, 37\%
(11\% of all reviews) and 44\% (11\% of all reviews) said they did,
respectively.
The ASE reviewers were asked what informed their guesses. The answers
were guessing based on paper topic (75 responses); obvious unblinding via reference to
previous work, dataset, or source code (31); having previously reviewed or read
a draft (21); or having seen a talk (3).
The results suggest that some deanonymization may be unavoidable. 
Some reviewers discovered GitHub repositories
or project websites while searching for related work to inform their reviews.
Some submissions represented clear extensions of or indicated
close familiarity with the authors' prior work.  However, there also exist
straightforward opportunities to improve
anonymization.  For example, community familiarity with anonymization, 
consistent norms, and clear guidelines could address the 
incidence of direct unblinding.
However, multiple times at the PC meetings, the PC chairs heard a PC member remark
about having been sure another PC member was a paper author, but
being wrong. Reviewers may be overconfident, and 
sometimes wrong, when they think they know an author through indirect unblinding.

\section{PC Chairs' Observations}
\label{sec:discussion}

After completing the process, the PC chairs of all three conferences
reflected on the successes and challenges of double-blind review. All PC
chairs were strongly supportive of continuing to use double-blind review in
the future. All felt that double-blind review mitigated effects of
(subconscious) bias, which is the primary goal of using double-blind
review. Some PC members also felt so, indicating anecdotally that they were more
confident that their 
reviews and decisions had less bias. One PC member remarked that double-blind
review is liberating, since it allows for evaluation without
concern about the impact on the careers of people they
know personally.

All PC chairs have arguments in support of their respective decisions on the
timing of revealing the authors (i.e., after review submission, before
PC meeting, or only for accepted papers).
The PLDI PC chair advocated strongly
for full double-blind, which enables rejected papers to be anonymously resubmitted to
other double-blind venues with common reviewers, addressing
one cause of deanonymization.
The ASE PC chairs observed that in a
couple of cases, revealing author identities helped to better understand a
paper's contribution and value.  The PLDI PC chair revealed author identities on
request, when deemed absolutely necessary to assess the paper. This happened
extremely rarely, and could provide the benefit observed by the ASE PC
chairs without sacrificing other benefits. 
That said, one PC member remarked that one
benefit of serving on a PC is learning who is working on what; full
anonymization eliminates learning the who, though still allows learning the
what.  

Overall, none of the PC chairs felt that the extra administrative
burden imposed by double-blind review was large.
The ASE PC chairs recruited two review process chairs to assist, and all felt
the effort required was reasonable. 
The OOPSLA PC chair noted the level of effort required to implement
double-blind review, including the management of conflicts of interest, was not
high.  He observed that it was critical to 
provide clear guidance to the authors on how to anonymize papers. 
(e.g., {\small \url{http://2016.splashcon.org/track/splash-2016-oopsla\#FAQ-on-Double-Blind-Reviewing}}).  
PLDI allowed authors to either anonymize artifacts (e.g., source code) or
to submit non-anonymized versions to the PC chair, who distributed to reviewers
when appropriate, on demand. The PC chair reported that this presented only a
trivial additional administrative burden. 

The primary source of additional administration in double-blind review is
conflict of interest management.  This task is simplified by conference
management software that  
straightforwardly allows authors and reviewers to declare conflicts based on
names and affiliations, and chairs to quickly cross-check declared
conflicts.  
ASE PC chairs worked with the CyberChairPro maintainer to support this task.
Neither ASE nor OOPSLA observed unanticipated conflicts discovered 
when author identities were revealed.
The PLDI PC chair managed conflicts of interest more creatively, creating a
script 
that validated author-declared conflicts by emailing PC members lists of
potentially-conflicted authors mixed with a random selection of other authors,
and asking the PC member to identify conflicts. The PC chair
examined asymmetrically declared conflicts and contacted authors regarding
their reasoning. This identified erroneous conflicts in rare instances. 
None of the PC chairs found identifying conflicts overly burdensome.
The PLDI PC chair reiterated that the burden of full double-blind reviewing
is well worth maintaining the process integrity throughout the entire
process, and for future resubmissions.

\section{Conclusions}
\label{sec:conclusions}

Data from ASE 2016, OOPSLA 2016, and PLDI 2016 suggest that, while
anonymization is imperfect, it is fairly effective. 
The PC chairs of all three conferences strongly support the continued use of double-blind
review, find it effective at mitigating (both conscious and subconscious) bias in reviewing, and judge the
extra administrative burden to be relatively minor and
well-worth the benefits. Technological advances and the now developed author
instructions reduce the burden. Having a dedicated organizational position to
support double-blind review can also help. The ASE and OOPSLA PC chairs point
out some benefits of revealing author identities mid-process, while the PLDI
PC chair argues some of those benefits can be preserved in a full
double-blind review process that only reveals the author identities of
accepted papers, while providing significant additional benefits, such as
mitigating bias throughout the entire process and
preserving author anonymity for rejected paper resubmissions.

\section{Acknowledgments}

Kathryn McKinley suggested an early draft of the reviewer questions used by
OOPSLA and PLDI.
This work is supported in part by the National Science
Foundation under grants 
CCF-1319688, 
CCF-1453474, 
CCF-1563797, 
CCF-1564162, 
and
CNS-1239498, 
and by the German Research Foundation (AP 206/6). 

\balance

\small
\bibliographystyle{plain}
\bibliography{blind}

\end{document}